\documentclass{www2010-submission}
\usepackage{url}
\usepackage{soul}
\begin{document}
\title{A Plan For Curating ``Obsolete Data or Resources''}

%\subtitle{[Extended Abstract]
%\titlenote{A full version of this paper is available as
%\textit{Author's Guide to Preparing ACM SIG Proceedings Using
%\LaTeX$2_\epsilon$\ and BibTeX} at
% You need the command \numberofauthors to handle the "boxing"
% and alignment of the authors under the title, and to add
% a section for authors number 4 through n.
%
% Up to the first three authors are aligned under the title;
% use the \alignauthor commands below to handle those names
% and affiliations. Add names, affiliations, addresses for
% additional authors as the argument to \additionalauthors;
% these will be set for you without further effort on your
% part as the last section in the body of your article BEFORE
% References or any Appendices.

\numberofauthors{1}
%
% Put no more than the first THREE authors in the \author command

% NOTE: All authors should be on the first page. For instructions
% for more than 3 authors, see:
% http://www.acm.org/sigs/pubs/proceed/sigfaq.htm#a18

\author{
%% e-mail address with \email.
\alignauthor Michael L. Nelson \\
       \affaddr{Old Dominion University} \\
       \affaddr{Norfolk, VA USA}\\
       \email{mln@cs.odu.edu}
}

\maketitle
\begin{abstract}
\sloppy
Our cultural discourse is increasingly carried in the web.  With the
initial emergence of the web many years ago, there was a period where
conventional mediums (e.g., music, movies, books, scholarly publications)
were primary and the web was a supplementary channel.  This has now
changed, where the web is often the primary channel, and other publishing
mechanisms, if present at all, supplement the web.  Unfortunately, the
technology for publishing information on the web always outstrips our
technology for preservation.  My concern is less that we will lose data
of known importance (e.g., scientific data, census data), but rather that
we will lose data that we do not yet know is important.  In this paper
I review some of the issues and, where appropriate, proposed solutions
for increasing the archivability of the web.
\fussy

\end{abstract}

% A category with only the three required fields
\category{H.3.7}{Digital Libraries}{}
%A category including the fourth, optional field follows...
%\category{D.2.8}{Software Engineering}{Metrics}[complexity measures,
%performance measures]

%\terms{Design, Experimentation, Standardization}

\keywords{Curation, Web Archiving, Memento}

\section{Who wants ``obsolete data''?}
\label{sec:introduction}

Perhaps the largest problem facing web archiving is that it remains 
at the fringes of the larger web community.  The most illustrative 
anecdote pertains to a web archiving paper we submitted to the 2010
WWW conference.  One of the reviews stated:

\begin{quote}
Is there (sic) any statistics to show that many or a good number of Web 
users would like to get obsolete data or resources? 
\end{quote}

This is just one reviewer, but the terminology used (``obsolete data or
resources'') succinctly captures the problem: web archiving is not widely
seen as a priority or even as in scope for a conference such as WWW.
Another common related misconception we have encountered is that the
Internet Archive has every copy of everything ever published on the web,
so preservation is a solved problem.  Despite the heroic efforts of the
Internet Archive, the reality is more grim: only 16\% of the resources
indexed by search engines are archived at least once in a public web
archive \cite{Ainsworth:2011:MWA:1998076.1998100}.   \\

While there are many specific challenges with regards to quality
criteria, tools, and metrics, the common thread goes back to
the fact that we, the web archiving community, have failed to
articulate clear, compelling use cases and demonstrate immediate
value for web preservation.  For too long web preservation has been
dominated by threats of future penalties, such as hoary stories
about file obsolescence that have not come true\footnote{David
Rosenthal has a series of convincing blog posts on this topic, see:
\url{http://blog.dshr.org/2010/09/reinforcing-my-point.html}}.  The lack
of a compelling use case for archives has relegated preservation to an
insurance-selling idiom, where uptake is unenthusiastic at best.\\

\section{I Blame Thompson and Ritchie}
\label{sec:inode}

The web has a poor notion of time, and it is getting worse instead
of better.  An early design document for the Web addressed the problem
of generic vs. specific resources \cite{tbl:generic}.  That document
identified three dimensions of genericity: time, language (e.g., English
vs. French), and representation (e.g., GIF vs. JPEG).  The latter two
dimensions were the basis for HTTP content negotiation as originally
defined in HTTP/1.1 \cite{rfc2068}.  Content negotiation allowed, for
example, GIF and JPEG resources to have unique URIs (i.e., specific
resources), but to be joined together with a third, generic resource
with its own URI.  When a client dereferences this generic URI, the
appropriate specific resource is selected based the client's preferences
for representations.  Content negotiation works similarly for language,
but content negotiation in the dimension of time was not part of the
original HTTP core technologies (the Memento project added content
negotiation in the dimension of time in 2009 \cite{nelson:memento:tr}).
One result of not having time as part of the core technologies is that
the web community's concept and expectations regarding time have not
become fully mature.\\

I believe the reason for this underdeveloped notion of time can be
traced to the tight historical integration of HTTP and Unix,
specifically the Unix filesystem.  Metadata about files in the Unix
filesystem is stored in ``inodes'', and the original description of the
Unix filesystem defined three notions of time to be stored in an inode:
file creation, last use, and last modification \cite{ritchie1974unix}.
However, at some early point the storage of the file creation time in
the inode was replaced with the last modification time of the inode
itself.  The result was that we could know the last modification and
access times of a file, but the creation time, a crucial part of
establishing provenance, was lost (most URIs contain semantics, and
creation time can be critical in establishing priority).  Although web
resources and Unix files are logically separate, in practice they were
tightly integrated during the formative years of the web, and so the
HTTP time semantics were limited by what could be provided by the Unix
inode.  For example, here is an HTTP response about a JPEG file:

\begin{scriptsize}
\begin{verbatim}
% curl -I cdn.loc.gov/images/img-head/logo-loc.png
HTTP/1.1 200 OK
Date: Sun, 19 Aug 2012 13:30:06 GMT
Server: Apache
Last-Modified: Fri, 03 Aug 2012 03:54:26 GMT
Content-Length: 1447
Connection: close
Content-Type: image/png
\end{verbatim}
\end{scriptsize}

In the above example, the server is expressing the response was sent on
August 19th, but the JPEG file itself was last modified on August 3rd.
Notable by its absence is the creation time: via the inode limitations,
we cannot know when this file was created.  It might have been created on
August 3rd or it might have been created at an earlier time, and being
unable to establish even this basic level of metadata is a severe
limitation for archiving and provenance.  Unfortunately, even the
limited semantics of last modified are becoming less frequent as more
resources are dynamically generated.  The example below is in response
for a dynamically generated home page:

\begin{scriptsize}
\begin{verbatim}
% curl -I www.digitalpreservation.gov/
HTTP/1.1 200 OK
Date: Sun, 19 Aug 2012 13:30:33 GMT
Server: Apache
X-Powered-By: PHP/5.2.8
Connection: close
Content-Type: text/html
\end{verbatim}
\end{scriptsize}

In the above example, there is the data of the response (August 19th),
but last modified times for dynamically generated representations are
not defined.  Dynamically generated resources make possible the web as
we know it today, but the net result is even fewer time semantics are
present in HTTP responses.  Evolving publishing technologies such as
personalization, Ajax, Flash, and streams\footnote{For example, see Anil
Dash's call to ``Stop Publishing Web Pages'' in favor of streams: 
\url{http://dashes.com/anil/2012/08/stop-publishing-web-pages.html}}
will only serve to make it more difficult to ascribe a creation time to
any particular web page.

\section{W\{h\}ither Archives?}
\label{sec:lack}

I maintain that the entire web community has a poor notion of time and
are trapped in the ``perpetual now''.  Because the lack of capability
has shaped our expectations, we never object when prior versions of
web pages are unavailable.  We tolerate temporal inconsistency in our
browsing, even 404 errors, in part because we do not know enough to
expect better.  Remember ``lost in hypertext'' \cite{elm1985getting,
conklin1987hypertext}?  That has been solved in part through better
navigation tools and design practices, but also in part due to increased
familiarity with the hypertext navigation metaphor.  Now imagine if
a temporal dimension was added for each page -- there would be much
confusion, but eventually tools, practices, and user awareness would
prevail.

\subsection{Archives Are Not Destinations}
\label{sec:destinations}

\begin{figure} 
\begin{center}
\includegraphics[scale=0.24]{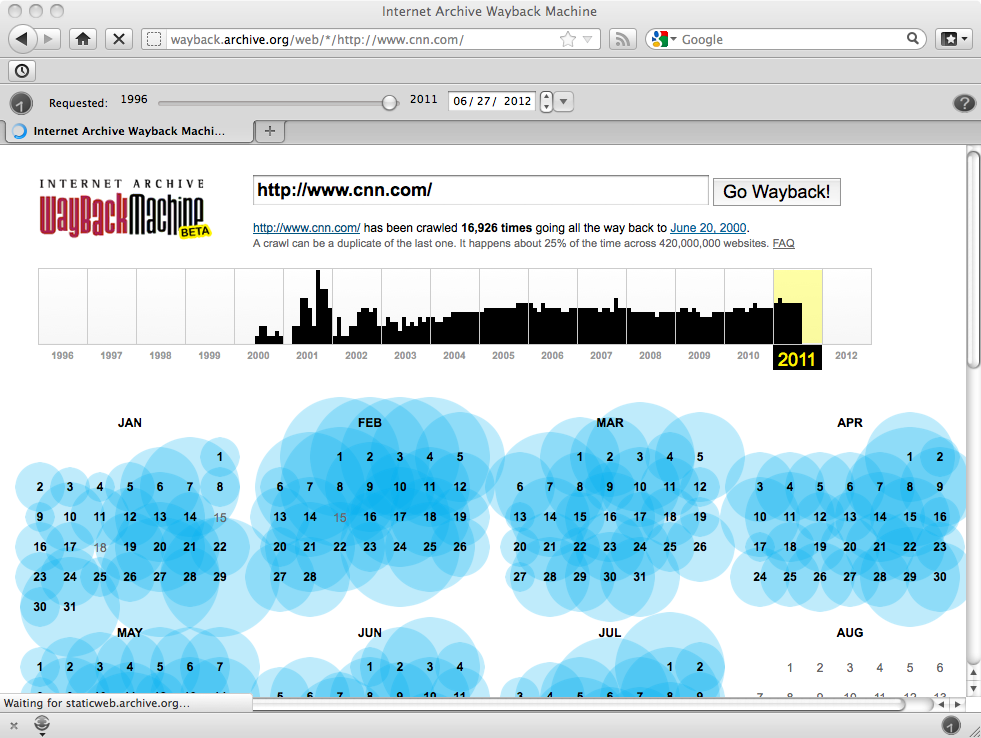}
\caption{All available versions of cnn.com at the Internet Archive.  This page is
not reachable from cnn.com.}
\label{fig:cnn-coffee-stain}
\end{center}
\end{figure}

The most fundamental problem is that we have designed web archives
as if they are destinations in themselves.  The motif of ``go to the
library/archive and spend an afternoon in the stacks'' has been replicated
in our web archives.  Figure \ref{fig:cnn-coffee-stain} shows the list 
of archived pages (or ``mementos'') for \url{cnn.com} at the Internet 
Archive.  If you want to browse the past versions of this news site, you
go to the archive and perform a browsing session within the archive,
and then return to the live web once you are done with your journey to the past. \\

In our experience, most web users do not know about the Internet Archive
or how to access it.  The Memento project has demonstrated a framework
for tighter integration of the past (i.e., archived) web and the current
web, but the tools exist as add-ons for both servers and clients and
have yet to reach mainstream acceptance, which will only arrive when
the archiving community can demonstrate a ``killer app'' that will cause
users to demand the functionality.

\subsection{Web Archiving Is Not Social}
\label{sec:social}

I am not sure what an archiving killer app would look like, but there
is a good chance it will be social.  People like to share links with
each other via Twitter, Facebook, Pinterest, et al.  However, with
the exception of Pinterest (which makes copies of ``pinned'' images)
this sharing is done by-reference and not by-value, exposing it to the
same link rot problems of common web pages (for example, we found 10\%
of the shared links about the Egyptian Revolution were lost after one
year \cite{TPDL2012:Losing}).  I am constantly surprised at the tasks that
people are willing to undertake if there is a social or gaming component
(i.e., ``games with a purpose''), yet I am unaware of any such activity
with a web preservation component.  Diigo (\url{diigo.com}) is a site
that provides social bookmarking services (similar to Delicious) with
an archiving component, but enthusiasm for social bookmarking seems to
be less than it once was.  \\

A web archiving application that could leverage the collection development
of Pinterest and the collaborative editing of Wikipedia and other wikis
would be a welcome development.  Archive-It (\url{archive-it.org})
is nearly such an application, but it is targeted for archiving and
librarian professionals, not as a general purpose social application.
Perhaps the legal challenges\footnote{A discussion of which is beyond
the scope of this paper; for a primer see \url{http://1.usa.gov/QgaUZO}}
of creating such collections would prevent the development of such an
application, but I would observe that early legal challenges about the
mechanics of HTTP and ``making copies'' were eventually overcome.

\subsection{Watchdog Archiving and Trust}
\label{sec:tds}

Perhaps a social web archiving activity that will grow to take
on a larger role is that of distributed, citizen watchdogs of
public figures and politicians.  For example, a supporter of
blogger Andrew Breitbart brought down Congressman Anthony Weiner by
zealously following and archiving Weiner's twitter feed\footnote{See
\url{http://en.wikipedia.org/wiki/Anthony_Weiner_sexting_scandal}}.
Most tweets are of arguably limited historical value, but this particular
tweet and the fact that it could not be fully redacted turned out to
have significant political and cultural implications.  \\

In another example, consultant and commentator Richard Grenell deleted
over 800 tweets after he was elevated to a senior position in the Romney
campaign in 2012\footnote{See: \url{http://huff.to/I6dpQo}}.  Presumably
Grenell's lesser status at the time did not warrant a corresponding
campaign to monitor and archive Grenell's twitter feed like there was
with Weiner's twitter feed.  Grenell's tweets most likely do not exist
outside of Twitter's own archives (and those they share with the Library
of Congress).\\

And what if someone did come forward with a correspondingly damning
tweet from Grenell, how could we verify it?  Aside from Weiner's
ultimate confession, was his tweet ever verified by an independent
third party?  And if so, how would we trust such a third party -- where
would the chain of trust terminate?  Could he not find a
technologically savvy staffer to fabricate evidence that contradicted
Breitbart's evidence (which is especially easy given the low level of
provenance regarding third-party archives)?  It is easy to envision a
market for a trusted, tamper-proof archive for tweets and other social
media so a person can \emph{deny} that they ever released an offending
tweet? \\

Our current approach to web archiving involves implicitly trusting
the Internet Archive and other public web archives as incorruptible.
Eventually the magnitude of scandals associated with web content will grow
to the point where less scrupulous web archives will be offered as proof.
A combination of trusted archives and citizen activism might form the 
basis for the first killer app for web archiving.  Instead of canvassing 
a neighborhood, volunteers can canvass/archive web pages.

\section{Wish List}
\label{sec:wish}

This section contains a personal wish list of features that would make 
archiving web pages much easier.  

\subsection{Machine-Readable Time Semantics}
\label{sec:machine}

We have moved beyond the limitations of the Unix filesystem and its inode,
so we should increase the time semantics in our HTTP transactions.
Unfortunately, this is not the case.  In the example below, when
dereferencing the URI of a specific tweet, twitter.com shows a last
modified time that matches the date the response was generated (this is
true for all responses, not just this one).  More importantly, Twitter has
a concept of time similar to ``Memento-Datetime'', which captures the time
a page was first observed on the web (see \cite{nelson2011memento} for
a discussion of how this differs from ``Last-Modified'').  Although this
date (June 27, 2012 in this example) is displayed in the HTML page and is
accessible to authenticated users via the Twitter API, the correct date
semantics are not presented, and the incorrect value for the last modified
time is presented instead.  This phenomenon is not unique to Twitter,
but Twitter makes for a good example due to its well-known nature.

\begin{scriptsize}
\begin{verbatim}
% curl -I twitter.com/machawk1/status/218015444496416768
HTTP/1.1 200 OK
Date: Mon, 20 Aug 2012 00:41:38 GMT
Content-Length: 85440
Last-Modified: Mon, 20 Aug 2012 00:41:38 GMT
Content-Type: text/html; charset=utf-8
Server: tfe
\end{verbatim}
\end{scriptsize}

\subsection{APIs for Archives}
\label{sec:apsi}

Talk to anyone who has built applications using archived web data and
they will have crawled and ``page scraped'' the archives at some point.
Page scraping puts an undue burden on the archive itself, is error prone,
and doesn't facilitate inter-archive interaction.  The Memento project
defines a simple, inter-archive HTTP access mechanism, but this is not
enough.  The Internet Archive's Wayback Machine software supports a simple
API for file upload and searching, but this API is not evolved like APIs
for services like Google, Twitter, and Facebook.  If we want archives to
be used in the current web programming idiom, we have to go beyond the
``afternoon in the stacks'' model (see section \ref{sec:destinations})
and provide fully-featured APIs.

\subsection{Impedance Matching}
\label{sec:travel}

The Internet Archive does not have full-text search on the main Wayback
Machine.  While this is a limitation, it is probably not as big a
limitation as many think, in part because it is not clear what we would
do with full-text search at this scale if we had it (cf. the discussion
in section \ref{sec:lack}).  The kinds of questions that scholars wish
to answer using web archives are of the form ``what role did the Tea
Party play in the 2010 mid-term elections?''  The kind of access we can
offer right now is ``this is what \url{cnn.com} looked like November 1,
2010.'' Adding full-text searching, while useful in some cases, would
not immediately help address the kinds of questions that scholars want
to ask.  An example of the kind of advanced analysis that needs to be
performed on web archives is entity tracking experiments of the LAWA
project \cite{Spaniol:2012:TEW:2187980.2188030}, in which entities (e.g.,
people, companies) can be tracked through time and different URIs.

\section{Conclusions}
\label{sec:conclusions}

I expect data of known value to be successfully curated and available
well into the future.  I am more concerned with our cultural record, 
with which we have made a Faustian bargain of increased volume and ease
of access (i.e., the web) at the expense of permanence and provenance
(i.e., paper).  We are stuck in the perpetual now and due to the initial
limitations of the Unix inode there, the notion of varying temporal
access to web pages is so unexpected that even web researchers need to
be convinced of the utility.\\

One problem is the limited design motif for web archives: destinations
that are wholly unconnected from their live web counterparts.  The related
problem is that we, as a community, have failed to envision and deliver
a ``killer app'' for web archiving.  Perhaps it is in a watchdog
role over public figures and institutions.  Or perhaps the emerging
field of personal digital preservation\footnote{See for example:
\url{http://www.personalarchiving.com/}} will energize the field and
increase what are often laissez-faire user expectations regarding
archiving \cite{Marshall07:Evaluating}. \\

I would like to see a more careful approach to specifying temporal 
semantics in common web services like Twitter.  Similarly, I expect
web archives to offer richer APIs for accessing their content, and 
to eventually offer the higher-level services, like entity tracking, 
that will assist scholars in using the \st{obsolete data or resources}
archives.  \\

%versioning like CMSs

%ACKNOWLEDGMENTS are optional
\section{Acknowledgments}
This work sponsored in part by the Library of Congress, NSF IIS-0643784
and IIS-1009392.

\bibliographystyle{abbrv}
\bibliography{mln}  % sigproc.bib is the name of the Bibliography in this case

\begin{thebibliography}{10}

\bibitem{Ainsworth:2011:MWA:1998076.1998100}
S.~G. Ainsworth, A.~Alsum, H.~SalahEldeen, M.~C. Weigle, and M.~L. Nelson.
\newblock How much of the web is archived?
\newblock In {\em Proceeding of the 11th annual international ACM/IEEE Joint
  Conference on Digital Libraries}, JCDL '11, 2011.

\bibitem{tbl:generic}
T.~{Berners-Lee}.
\newblock Web architecture: Generic resources.
\newblock http://www.w3.org/DesignIssues/Generic.html, 1996.

\bibitem{conklin1987hypertext}
J.~Conklin.
\newblock Hypertext: A survey and introduction.
\newblock {\em IEEE Computer}, 20(9):17--41, 1987.

\bibitem{elm1985getting}
W.~Elm and D.~Woods.
\newblock Getting lost: A case study in interface design.
\newblock In {\em Proceedings of the Human Factors and Ergonomics Society
  Annual Meeting}, volume~29, pages 927--929, 1985.

\bibitem{rfc2068}
R.~Fielding, J.~Gettys, J.~Mogul, H.~Frystyk, and T.~Berners-Lee.
\newblock {Hypertex Transfer Protocol -- HTTP/1.1, Internet RFC-2068}, 1997.

\bibitem{Marshall07:Evaluating}
C.~Marshall, F.~McCown, and M.~L. Nelson.
\newblock Evaluating personal archiving strategies for {Internet-based} in
  formation.
\newblock In {\em Proceedings of IS\&T Archiving 2007}, pages 151--156, May
  2007.

\bibitem{nelson2011memento}
M.~L. Nelson.
\newblock {Memento-Datetime is not Last-Modified}.
\newblock
  \url{http://ws-dl.blogspot.com/2010/11/2010-11-05-memento-datetime-is-not-last.html},
  2011.

\bibitem{ritchie1974unix}
D.~Ritchie and K.~Thompson.
\newblock The {UNIX} time-sharing system.
\newblock {\em Communications of the ACM}, 17(7):365--375, 1974.

\bibitem{TPDL2012:Losing}
H.~M. SalahEldeen and M.~L. Nelson.
\newblock Losing my revolution: How much social media content has been lost?
\newblock In {\em TPDL}, 2012.

\bibitem{Spaniol:2012:TEW:2187980.2188030}
M.~Spaniol and G.~Weikum.
\newblock Tracking entities in web archives: the {LAWA} project.
\newblock In {\em Proceedings of the 21st international conference companion on
  World Wide Web}, WWW '12 Companion, 2012.

\bibitem{nelson:memento:tr}
H.~{Van de Sompel}, M.~L. Nelson, R.~Sanderson, L.~L. Balakireva, S.~Ainsworth,
  and H.~Shankar.
\newblock {Memento: Time Travel for the Web}.
\newblock Technical Report arXiv:0911.1112, 2009.

\end{thebibliography}
% ACM needs 'a single self-contained file'!
\balancecolumns
\balancecolumns % GM July 2000
% That's all folks!
\end{document}